\documentclass[aps,preprint]{revtex4}
\usepackage[dvips]{graphics,graphicx}
\begin{document}
\title{Energetically constrained co-tunneling of cold atoms }
\author{Andrey R. Kolovsky$^{1,2}$}
\affiliation{$^1$Kirensky Institute of Physics, 660036 Krasnoyarsk, Russia}
\affiliation{$^2$Siberian Federal University, 660041 Krasnoyarsk, Russia}
\author{Julia Link$^{3}$}
\author{Sandro Wimberger$^{3}$}
\affiliation{Institut f\"ur Theoretische Physik, Universit\"at Heidelberg, 69120 Heidelberg, Germany$^3$}
\date{\today}

\begin{abstract}
We study under-barrier tunneling for a pair of energetically bound bosonic atoms in an optical lattice with a barrier. We identify conditions under which this exotic molecule tunnels as a point particle with the coordinate given by the bound pair center of mass and discuss the atomic co-tunneling beyond this regime. In particular, we quantitatively analyze resonantly enhanced co-tunneling, where two interacting atoms penetrate the barrier with higher probability than a single atom.
\end{abstract}

\maketitle

\section{Introduction}
\label{sec1}

The phenomenon of under-barrier tunneling is one of the most exciting predictions of quantum mechanics, which does not fit the classical picture of the world. As a bright example of a pure quantum effect, it is considered in all text books on quantum mechanics, which might make an impression that under-barrier tunneling has been exhaustively studied already in early days of quantum mechanics. However, the theoretical description of the under-barrier tunneling is simple only for a point particle in one dimension. If we have a composite object, the problem of tunneling becomes very involved. Systematically this problem was first addressed in nuclear physics (see \cite{Zakh96} and references therein).  It was found that the tunneling probability for the composite object may considerably differ from that for the point particle of the same mass.

In the present work we revisit the problem of the under-barrier tunneling for a composite object which has become experimentally available only recently -- the pair of interacting bosonic atoms in an optical lattice,  where two atoms stay close to each other due to the energy constrain \cite{Wink06}.  Note that such atomic pairs exist for both attractive and repulsive inter-atomic interactions if one satisfies the necessary condition that the interaction energy $|U|$ exceeds the single-atom tunneling energy $J$. Following Ref.~\cite{Wink06} we shall refer to this exotic two-atom molecule as the bound pair. It admits a simple yet rigorous description \cite{Scot94,Vali08,Nyga08,Java10}, which greatly facilitates theoretical studies of different aspects of the composite object tunneling. 

Our other motivation for studying the under-barrier tunneling of the bound pair are problems of macroscopic tunneling of a Bose-Einstein condensate of cold atoms \cite{Albi05,Paul07,Deke10,Glik11}. In particular, the recent paper \cite{Glik11} analyzes numerically the under-barrier tunneling of a bright soliton consisting of $N\gg1$ condensed atoms. Since the bound pair can be considered as the bright soliton with $N=2$ atoms \cite{Scot94,Flac08}, rigorous analysis of the bound-pair tunneling may shed additional light on the phenomenon of macroscopic tunneling.

The structure of the paper is as follows. In Sec.~\ref{sec2} we recall the main results on eigenstates of the bound pair in the absence of an external potential and introduce a simple two-state model, which suffices   to describe the mobility of the bound pair. Sec.~\ref{sec3} is devoted to the under-barrier tunneling within the framework of the two-state model. We identify the conditions under which the bound pair tunnels as the point particle with the coordinate given by the center of mass of the bound pair and uncover the effect of resonant tunneling, which is entirely due to internal degrees of freedom of a composite object. The main drawback of the two-state model is that it neglects the dissociation process where the barrier breaks the pair into two unbound atoms.  For this reason, in Sec.~\ref{sec4} we simulate the tunneling process numerically without using any approximations.  We summarize our findings in the concluding Sec.~\ref{sec6}.

\section{Two-state and one-state models}
\label{sec2}

\begin{figure}[b]
\center
\includegraphics[width=10cm,clip]{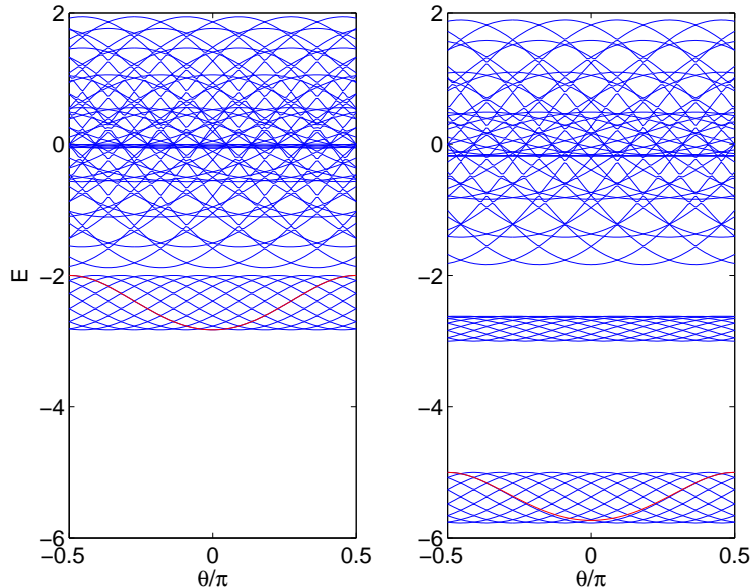}
\caption{The band spectrum of the system (\ref{1}) for $J=1$ and $(U_0,U_1)=-(2,0)$ (left panel) and $(U_0,U_1)=-(5,3)$ (right panel). The lattice comprises 11 sites with periodic boundary conditions. The red lines are Eq.~(\ref{4b}) and (\ref{4}), respectively.}
\label{Gfig1}
\end{figure}

As mentioned in the introductory section, the strongly interacting bosons in a lattice form bound pairs, where two bosons occupy the same site. Such a pair can move across the lattice by virtually  breaking the bond \cite{remark2} and, thus, is a composite object with well defined kinetic energy. The dispersion relation $E(\kappa)$ for the bound pair can be easily calculated numerically by diagonalizing the Hamiltonian of the Bose-Hubbard model [see Eq.~(\ref{1}) below] with $N=2$ particles. For the purpose of future references Fig.~\ref{Gfig1}(a) shows the result of this diagonalization for a lattice comprising $11$ sites,  where we additionally parametrize the Bose-Hubbard Hamiltonian by the Peierls phase $\theta$: $\hat{a}^\dag_{l+1}\hat{a}_l \rightarrow \hat{a}^\dag_{l+1}\hat{a}_l\exp(i\theta)$. In Fig.~\ref{Gfig1}(a) the bound pair of two bosons is associated with the lower band, while the upper band is the spectrum of two hard-core bosons. The problem can be also solved analytically, either exactly or by using a perturbative approach. In the rest of this section we  discuss the dispersion relation $E(\kappa)$ and the eigenstates of the bound pair within the perturbative approach, which better fits our aims of studying the tunneling process.

To facilitate the theoretical analysis it is convenient  to consider the Bose-Hubbard model which also includes interactions in neighboring sites:
\begin{equation}
\label{1}
\widehat{H}_{BH}=
-\frac{J}{2}\sum_{l} \left( \hat{a}^\dag_{l+1}\hat{a}_l +h.c.\right)
  +\frac{U_0}{2}\sum_{l} \hat{n}_l(\hat{n}_l-1)   
  + U_1\sum_l  \hat{n}_{l+1} \hat{n}_l   \;.
\end{equation}
Inclusion of the latter term explicitly introduces excited states of the bound pair, see Fig.~\ref{Gfig1}(b). We note that we do not assign the interaction constant $U_1$ any physical meaning \cite{Wang10} and let eventually $U_1$ tend to zero. The perturbative approach to the energy spectrum of the bound pair essentially amounts truncation the Hilbert space of the operator (\ref{1}) to the subspace which includes only the Fock states where two bosons occupy either the same site or two neighboring sites. Then, denoting by $\Psi_l^{(1)}$ the probability amplitude to find two bosons at the site $l$ and by $\Psi_l^{(2)}$ to find them at sites $l$ and $l+1$, the eigenvalue equation for the bound pair takes the form  $(\widehat{H}_0 \Psi)_l
=E \Psi_l$, where
\begin{equation}
\label{2}
(\widehat{H}_0 \Psi)_l =-\frac{\sqrt{2}J}{2}\left[ 
\left(\begin{array}{cc} 0&0\\1&0 \end{array}\right) \Psi_{l-1} 
+\left(\begin{array}{cc} 0&1\\1&0 \end{array}\right) \Psi_{l} 
+\left(\begin{array}{cc} 0&1\\0&0 \end{array}\right) \Psi_{l+1}
\right]
+\left(\begin{array}{cc} U_1&0\\0&U_0 \end{array}\right) \Psi_{l} \;.
\end{equation}
The solutions of this eigenvalue problem are plane waves, $\Psi_l={\bf C}(\kappa)e^{i\kappa l}$, where the vector ${\bf C}(\kappa)$ satisfies the following $2\times2$ eigenvalue equation:  
\begin{equation}
\label{3}
\left(\begin{array}{cc}
\Delta&-J\left(1+e^{i\kappa}\right)/\sqrt{2} \\
-J\left(1+e^{-i\kappa}\right)/\sqrt{2}&0
\end{array}\right) {\bf C}=E{\bf C} 
\;, \quad \Delta=|U_0-U_1| \;. 
\end{equation}
From (\ref{3}) we  have
\begin{equation}
\label{4}
E(\kappa)=U_0+\frac{\Delta}{2} \pm \sqrt{ \left(\frac{\Delta}{2}\right)^2 +
2J^2\cos^2\left(\frac{\kappa}{2}\right) } \;.
\end{equation}
In what follows we refer to Eqs.~(\ref{2}-\ref{4}) as the two-state model.

It is worth stressing that the two-state model provides only an approximation to the exact eigenstates of the bound pair. How good this approximation is depends on the system parameters. In general, the two-state model is a good approximation for both the ground and excited bands if $|U_1|,|U_0|\gg J$ and $\Delta\ll |U_1|,|U_0|$. If $|U_1|$ is decreased, yet $|U_0|\gg J$, it is still a reasonable approximation for the ground band [minus sign in Eq.~(\ref{4})]. This also includes the case $U_1=0$ where the excited band is `dissolved' into the spectrum of unbound bosons. In this case, according to our numerical analysis, $|U_0|$ should be at least twice as large as $J$. Then the admixture of the truncated Fock states (i.e., those belonging to the truncated subspace of the Hilbert space) to the exact ground state of the bound pair does not exceed 5 percents. For smaller $|U_0|$ the contribution of these Fock states cannot be neglected  and one has to diagonalize the Hamiltonian (\ref{1}) in the whole Hilbert space \cite{Vali08}, which results in the dispersion relation  
\begin{equation}
\label{4b}
E(\kappa)=-\sqrt{ U_0^2 + 4J^2\cos^2\left(\kappa/2\right) } \;.
\end{equation}
In Fig.~\ref{Gfig1} we plot the analytical results (\ref{4b}) and (\ref{4}) by the red lines.

If the band gap $\Delta\gg J$ the problem can be simplified further, resulting in the one-state model. The procedure goes as follows. First we restrict ourselves to the ground band and introduce the Wannier states $\Phi_l$ of the bound pair by integrating its translationally invariant eigenstate $\Psi_\kappa$ over the quasimomentum in the first Brillouin zone:
\begin{equation}
\label{5}
\Phi_l=\int_{-\pi}^{\pi}  \Psi_\kappa e^{-i\kappa l} {\rm d}\kappa 
=\int_{-\pi}^{\pi} {\bf C}^{(-)}(\kappa)e^{i(l'-l)\kappa}  {\rm d}\kappa \;.
\end{equation}
The Wannier states (\ref{5}) are localized functions with the center of gravity at the site $l$. (For example, for $J=1$, $U_0=-4$ and $U_1=0$ we have $|\Psi_0\rangle \approx 0.157 |\ldots,0,1,1,0,0,\ldots\rangle +0.975 |\ldots,0,0,2,0,0,\ldots\rangle + 0.157 |\ldots,0,0,1,1,0\ldots,\rangle$.)  Next we calculate matrix elements of the Hamiltonian (\ref{2}) for the Wannier states seperated by $m$ sites:
$I_m=\langle \Phi_{l+m}|\widehat{H}_0|\Phi_l\rangle$.
%
This way we obtain the effective Hamiltonian where the bound pair is considered as a point particle:
\begin{equation}
\label{7}
\widehat{H}_{eff}=
-\frac{1}{2}\sum_m I_m\sum_l \left( \hat{b}^\dag_{l+m}\hat{b}_l
+h.c.\right)   \;.
\end{equation}
In the limit $\Delta\gg J$ we have $I_1=2J^2/\Delta$ and one can safely neglect the next to neighboring hopping. Obviously, this situation corresponds to the case where the dispersion relations (\ref{4},\ref{4b}) are approximated by the cosine function (which in practice requires $\Delta>4J$).

\section{Under-barrier tunneling for the two-state model}
\label{sec3}

The one-state model (\ref{7}) introduced in the previous section gives us the reference frame in studying the under-barrier tunneling of the bound pair. To be certain we shall consider a Gaussian barrier, $\epsilon_l=V\exp(-l^2/2\sigma^2)$, and a plane wave coming from minus infinity. Since the Gaussian barrier is well localized within the finite interval  $|l|< L\sim\sigma$ we can find the tunneling probability by using, for example, the transfer matrix method (see Appendix). Alternatively, one finds the tunneling probability by simulating the scattering process for a localized wave packet on the basis of the time-dependent Schr\"odinder equation:
\begin{equation}
\label{9}
i\partial_t\psi_l=-\frac{1}{2}\sum_m I_m(\psi_{l+m} + \psi_{l-m}) +
2\epsilon_l\psi_l \;,\quad
\epsilon_l=V\exp(-l^2/2\sigma^2) \;.
\end{equation}
As the initial conditions for (\ref{9}) it is convenient to choose a wide Gaussian with the given group velocity, $\psi_l(t=0)=G(l-l_0)\exp(i\kappa l)$. If the width of this initial packet is large enough, the result of time-dependent simulations practically coincides with that obtained on the basis of the  stationary  Schr\"odinder equation. 

A remark concerning the sign of the parameter $V$ is in order. In what follows we consider both positive and negative $V$, i.e., potential barriers and wells. In fact, for a particle in a lattice the notions of `barrier' and  `well' are equivalent  to each other to some extent. This becomes especially clear in the case of neighboring hopping ($I_m=I_1\delta_{m,1}$), where scattering of the plane wave with the quasimomentum $\kappa$ on the well ($V<0$) is equivalent to scattering of the plane wave with the quasimomentum $\kappa'=\pi-\kappa$ on the barrier ($V>0$). [Note that for $\kappa=\pi/2$ this implies a symmetric function $P_t(V)$ for the tunneling probability.] Also, by considering both positive and negative $V$ we cover the case of repulsive interactions as well, with the obvious substitution $V\rightarrow -V$ when the sign of interaction constants is changed.
\begin{figure}
\center
\includegraphics[width=10cm,clip]{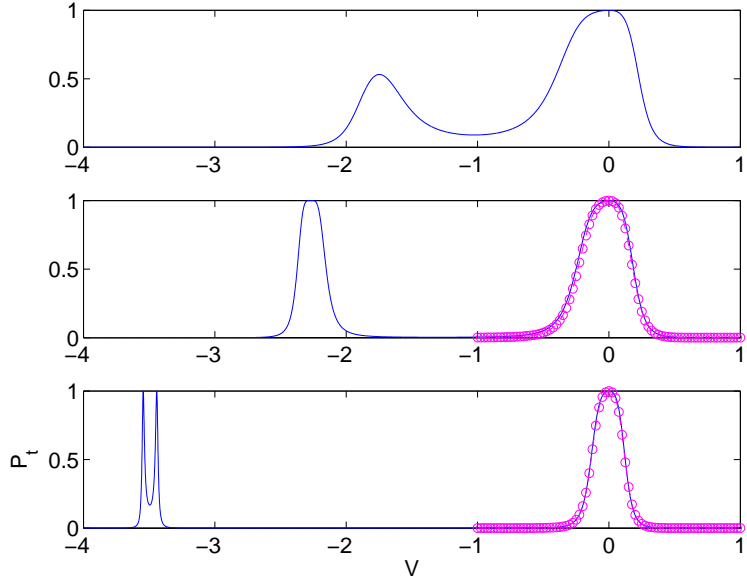}
\caption{Two-state model: Tunneling probability as the function of $V$ for fixed $\kappa=\pi/2$ and $\Delta=1,2,4$. The scattering potential is $\epsilon_l=V\exp(-l^2/2\sigma^2)$, $\sigma=0.65$. The open circles are predictions of the one-state model.}
\label{Gfig2}
\end{figure}

All said above about the one-state model is equally applied to the two-state model. Here, instead of (\ref{9}), one deals with the Schr\"odinder equation 
\begin{equation}
\label{10}
i\partial_t\Psi_l=(\widehat{H}_0 \Psi)_l 
+\left(\begin{array}{cc} \epsilon_l+\epsilon_{l+1}&0\\0&2\epsilon_l
\end{array}\right) \Psi_{l} \;,
\end{equation}
where $\Psi_l$ is a two-component vector and the Hamiltonian $\widehat{H}_0$ is defined in Eq.~(\ref{2}). Our particular interest is the scattering of a plane wave transmitting in the ground energy band.  As an example, Fig.~\ref{Gfig2} shows the tunneling probability for the plane wave with $\kappa=\pi/2$ for 3 different values of the parameter $\Delta$. In the panes (b,c) of this figure we also plotted the tunneling probability obtained on the basis of the one-state model.
\begin{figure}
\center
\includegraphics[width=13cm,clip]{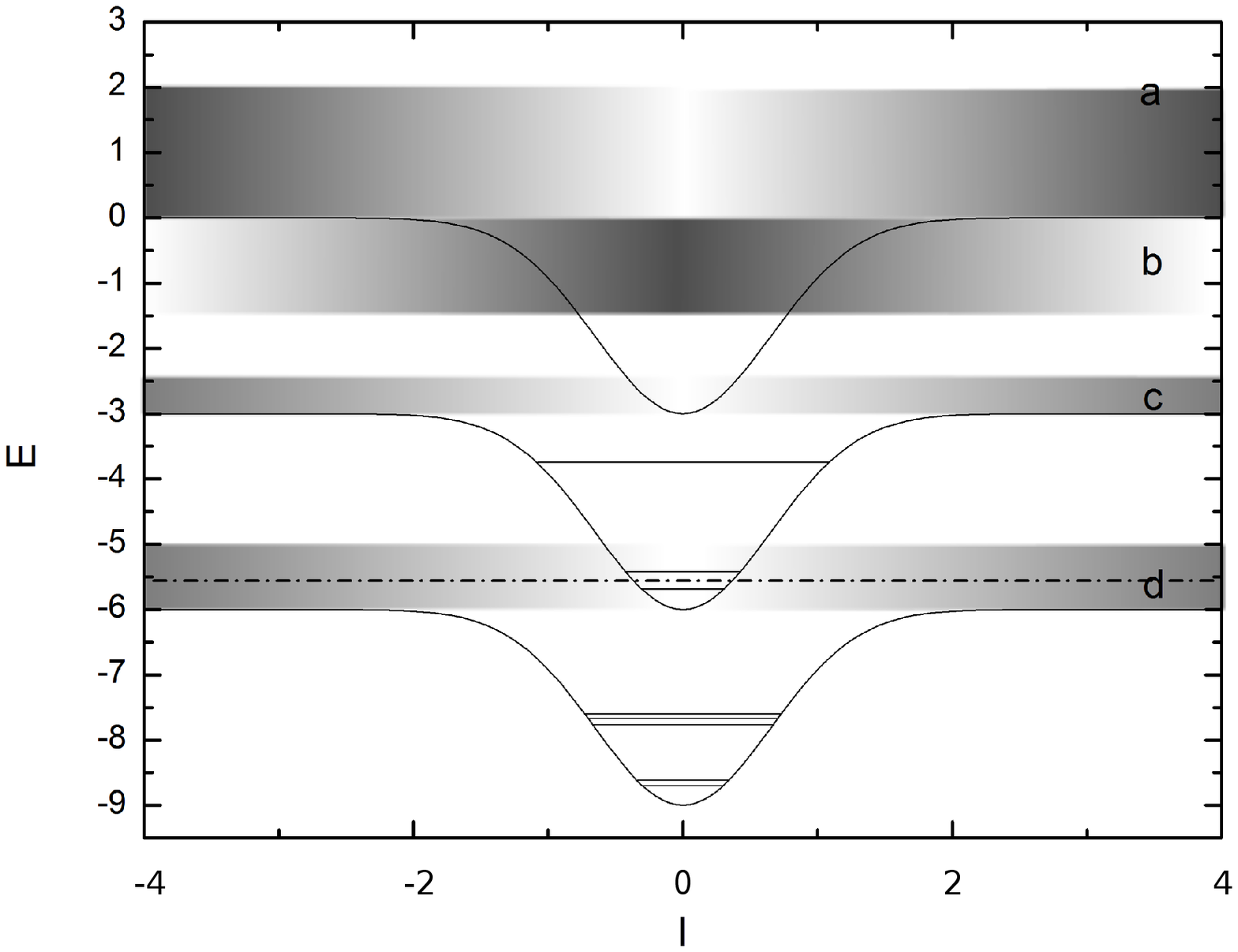}
\caption{Pictorial presentation of the resonant tunneling. The letters (d-c) labels the ground and excited energy bands of the bound pair, (b) is the energy band of two bosons with one of them captured in the well, and (a) is for unbounded bosons.}
\label{Gfig0}
\end{figure}
\begin{figure}
\center
\includegraphics[width=10cm,clip]{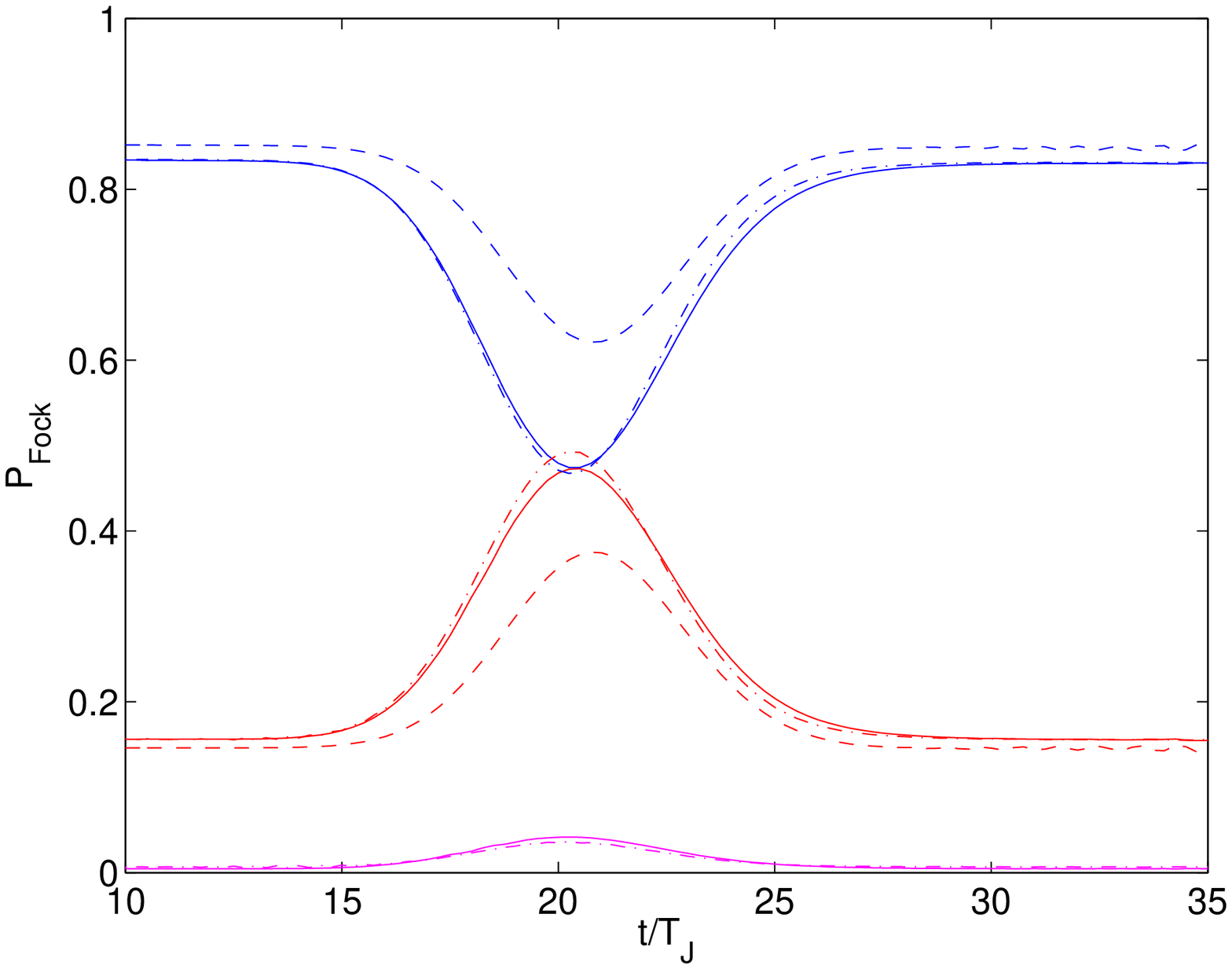}
\caption{The total occupation probability of Fock states with two bosons in the same site (upper curves), Fock states with two bosons in neighboring sites (middle curves), and Fock states with two bosons separated by empty sites (lower curves). The dashed, dash-doted, and solid lines distinguish between the two-state model, the three-state model and the original system. Parameters are $J=1$, $(U_0,U_1)=-(4,2)$ (hence $\Delta=2$), $\sigma=0.65$, and $V=-2.2$.}
\label{Gfig7}
\end{figure}

A remarkable prediction of the two-state model as compared to the one-state model is the appearance of narrow transparency windows for negative $V$. Usually such windows are associated with resonant tunneling in multi-barrier structures. In our case (single barrier or well) we meet a different type of resonant tunneling, where the bound pair tunnels through the upper band or, more precisely, through a localized state of the bound pair in the excited state, as it is pictorially shown in Fig.~\ref{Gfig0}. This interpretation of the enhanced tunneling is strongly supported by numerical simulations of the wave-packet dynamics, where we observe a temporal population of the upper band when the packet passes through the well. This is shown by the dashed line in Fig.~\ref{Gfig7}, where the upper curve is the total occupation probability of Fock states with two bosons in the same site and the middle curve is the total occupation probability of Fock states with two bosons in the neighboring sites. 

Number and widths of resonances seen in $P_t(V)$ crucially depend on the system parameters, in particular, on the width $\sigma$ of the potential well. If $\sigma$ is increased, we observe more resonances and they are narrower. The decrease of $\Delta$ makes resonances wider. It should be also mentioned that, for the currently considered Gaussian potential, the resonances appear in pairs, as seen in Fig.~\ref{Gfig2}(c), and the pair can merge into the single wide resonance, as it is the case depicted in Fig.~\ref{Gfig2}(a-b).

\section{The full system}
\label{sec4}

\begin{figure}
\center
\includegraphics[width=8cm,clip]{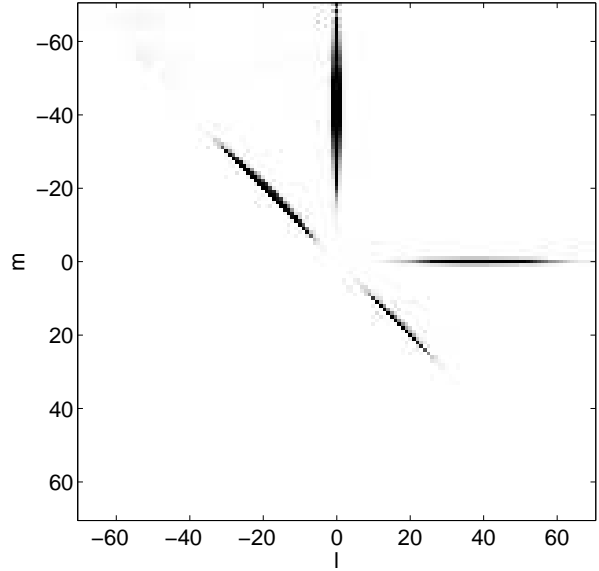}
\caption{Result of numerical simulations for the wave-packet dynamics of the original system for $J=1$, $U_0=-2$, $U_1=0$, $\sigma=0.65$, $V=-2$. The figure encodes (by using the gray-scaled mapping) the probability to find two bosons at sites $l$ and $m$.}
\label{Gfig3}
\end{figure}

Next we discuss the degree of validity of the two-state model. Indeed, the two-state model neglects the coupling to the truncated Fock states, which are associated with unbound bosons. If this coupling is strong (as in the case $U_1\approx 0$, where the upper band of the two-state model is embedded into the energy band of hard-core bosons) it may essentially affect the tunneling process and even open new scattering channels where the bound pair dissociates. For this reason we simulate the tunneling process on the basis of the Bose-Hubbard model (\ref{1}), i.e., without using any approximations. In this numerical experiment we propagate the wide Gaussian packet constructed from eigenstates of the bound pair for a time approximately twice longer than required for the packet to hit the potential barrier. Fig.~\ref{Gfig3} shows a typical result for $U_0=-2$, $U_1=0$ and $V =-2$. The figure depicts probabilities to find the system in the Fock state with one boson at the site $l$ and the other one at the site $m$. (Note that $m\ge l$ in the considered case of identical particles.)  It is seen in Fig.~\ref{Gfig3} that the initial packet splits into four packets where two of them, which are located at the main diagonal, are associated with the bound pair and the other two are the dissociated pair with one boson staying in the potential well. (For the chosen parameter the dissociation is energetically allowed because the boson in the well accumulates almost the whole bound energy.)  Summing up probabilities for this four  packets, which are well separated in the Fock space, we find the tunneling, reflection, and dissociation probabilities.
\begin{figure}
\center
\includegraphics[width=12cm,clip]{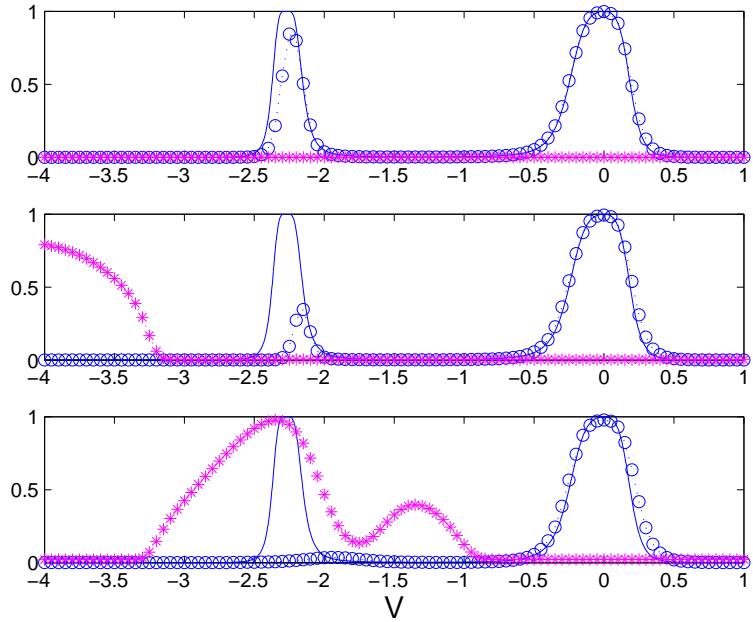}
\caption{Tunneling (open circles) and dissociation (asterisks) probabilities vs. $V$. The system parameters are $\kappa=\pi/2$ $J=1$, $\sigma=0.65$, and  $(U_0,U_1)=-(8,6)$ (top), $-(4,2)$ (middle), and $-(2,0)$ (bottom). The solid line is the prediction of the two-state model.}
\label{Gfig4}
\end{figure}

To systematically study the effect of the Hilbert space truncation (i.e., the effect of unbound bosons) we fixed the parameter $\Delta=|U_0- U_1|$ and vary the interaction energy $U_1$, where the limiting case corresponds to $U_1=0$. The parameter $\kappa$, which defines the group velocity of the incoming wave packet, is fixed to $\kappa=\pi/2$ and the parameter $\sigma$, which defines the width of the potential barrier, is $\sigma=0.65$ (then the scattering potential essentially comprises 3 lattice sites). The results of our numerical simulations are depicted in Fig.~\ref{Gfig4} by symbols, which are connected by the dotted line to guide eye. Open circles show the tunneling probability  and asterisks the dissociation probability. (The reflection probability, which is obviously given by $P_r=1-P_t-P_d$, is not shown.)  By inspection of the numerical data we can  draw the following conclusions: 
(i) There are practically no deviations from the predictions of the one- and two-state models for the main transparency window around $V=0$.  Thus, here the bound pair tunnels just like a point particle;
(ii) The system exhibits  resonant tunneling at $V\approx \Delta$, as predicted by the two-state model. This is a clear manifestation of complex structure of the tunneling object;
(iii) As compared to the two-state model the resonant tunneling is suppressed. It is interesting to note that the tunneling is suppressed independent of whether the dissociation channel is open or closed;
(iv) If the dissociation channel is open, we observe strong back action of the resonant tunneling on the dissociation process, which manifests itself in a local deep in $P_d=P_d(V)$.

Let us discuss the suppression of the resonant tunneling in some more details. We found this suppression to be fairly reproduced  if the Hilbert space of the two-state model is enlarged by including the Fock states, where two bosons are separated by one empty site. In the other words, instead of the two-state model one considers a three-state model. The dash-dotted lines  in Fig.~\ref{Gfig7} show population dynamics for the considered three families of Fock states. It is seen that it is practically coincides with that for the full system. Note that the third state remains practically unpopulated during the tunneling process. Nevertheless, the presence of this third state appears to be important.  This statement is also supported by the transfer matrix analysis, where the inclusion of this state considerably modifies  the resonant tunneling (see Appendix).

\section{Conclusions}
\label{sec6}

We studied the tunneling of an interactively bound pair of two bosons in 1D lattice through a narrow potential barrier/well. This system is, perhaps, the simplest composite object which can be created in the laboratory. We address the question under which conditions this composite object tunnels like  a point particle. Loosely speaking, these conditions amount to the requirement that the microscopic interaction constant $U_0$ entering the Bose-Hubbard Hamiltonian is larger than the hight of the potential barrier (the case of repulsive interactions) or the depth of the potential well (attractive interactions). If this condition is satisfied, the pair can be considered as a point particle. Note that the effective Hamiltonian for this `point particle' contains next to neighboring hopping, which is absent in the original Bose-Hubbard Hamiltonian. This takes into account the finite size of our composite object when it is treated as a point object \cite{Zakh96,Soko64}.

If the above condition is violated, one meets two phenomena which are entirely due to internal degrees of freedom of the composite object. These are  (i) the resonant tunneling and (ii) the dissociation.  Under conditions of resonant tunneling the bound pair can tunnel through the barrier, which the single boson  cannot penetrate. (Here we refer to the case of repulsive interactions.) Assuming the shape of the barrier to be fixed, resonant tunneling takes place in a rather small region of the parameter space  spanned by the quasimomentum $\kappa$ of the incoming plane wave (or group velocity of the incoming wave packet)  and the height $V$ of the potential barrier.  On the contrary, the parameter region where the external potential breaks the pair by capturing one boson at the barrier is relatively large. For a generic form of the external potential and $U_1=0$ the region of resonant tunneling is usually embedded into the dissociation region and, thus, the resonant tunneling and dissociation coexist. We observed strong mutual  influence of these processes which results in suppression of  both the tunneling and dissociation.

In the present work we calculated the tunneling and dissociation probabilities by simulating the wave-packet dynamics of the bound pair \cite{remark3} where, as the physical object, we had in mind ultra-cold atoms in the 1D optical lattice with the scattering potential created by additional laser beam crossing the lattice at the right angle. For this reason we considered a Gaussian shape of the potential barrier/well. It seems unlikely that one can obtain a compact analytical expression for the tunneling and dissociation amplitudes for this scattering potential. However, there are good prospects in developing the analytic theory for particular shapes of the barrier/well, which includes the impurity-like potential $\epsilon_l=V\delta_{l,0}$ and the box-like potential  $\epsilon_l=V(\delta_{l,0}+\delta_{l,1})$.  The advantage of the latter potential is that the well size and the size of the bound pair in the excited state matches exactly and, thus, the resonant tunneling and dissociation regions do not overlap.


\section{Appendix}

The transfer matrix propagates the plane wave solution $\psi_l=\exp(i\kappa l)$ from the asymptotic region $l<-L$ to the asymptotic region $l>L$, where the wave function is given by the superposition of two plane waves, $\psi_l=a\exp(i\kappa l)+b\exp(-i\kappa l)$. Then the tunneling amplitude is given by $t=1/b^*$. The latter relation follows from the equation
\begin{equation}
\label{b0}
\left(\begin{array}{c}
0\\ b
\end{array} \right)=S 
\left(\begin{array}{c}
1\\ a
\end{array} \right) \;, \quad
S=\left(
\begin{array}{rr}
r& -t^*\\
t &r^*
\end{array}
\right) \;,
\end{equation}
where $S$ is the scattering matrix. The relation $t=1/b^*$ is valid independent of whether the wave function is scalar or vector, although the explicit form of the transfer matrix is different. 

We begin with the one-state model (\ref{7}) where we assume $I_m=I\delta_{m,1}$  for simplicity:
\begin{equation}
\label{b1}
-\frac{I}{2}(\psi_{l+1} + \psi_{l-1}) + 2\epsilon_l\psi_l = E\psi_l \;, \quad 
E(\kappa)=-I\cos \kappa \;.
\end{equation}
It immediately follows from (\ref{b1}) that the plane wave can be propagated as
\begin{equation}
\label{b2}
\left(
\begin{array}{c}
\psi_{l+1}\\ \psi_{l}
\end{array}
\right)=T_l \left(
\begin{array}{c}
\psi_{l}\\ \psi_{l-1}
\end{array}
\right) \;,
\end{equation}
where
\begin{equation}
\label{b3}
T_l=\left(
\begin{array}{cc}
2(2\epsilon_l -E)/I & -1\\
1 & 0
\end{array}
\right) \;,
\end{equation}
is the one-step transfer matrix. It is worth noting that the matrix (\ref{b3}) is unitary, with the eigenvalues lying on the unit circle.

The case of the two-state model is more involved. Here Eq.~(\ref{b2}) takes the form
\begin{equation}
\label{b4}
\left(
\begin{array}{c}
\Psi_{l+1}^{(2)} \\ \Psi_{l+1}^{(1)}
\end{array}
\right)=T_l \left(
\begin{array}{c}
\Psi_{l}^{(2)}\\ \Psi_{l}^{(1)}
\end{array}
\right) \;,
\end{equation}
and the transfer matrix is given by
\begin{equation}
\label{b5}
T_l=\left(
\begin{array}{cc}
(ab-1)& -b\\
a & -1
\end{array}
\right) \;,
\end{equation}
where $a=\sqrt{2}(\Delta+\epsilon_l +\epsilon_{l+1} -E)/J$ and $b=\sqrt{2}(2\epsilon_{l+1}-E)/J$. The dispersion relation $E=E(\kappa)$ entering equation (\ref{b5}) is obtained by solving the Schr\"odinder equation in the absence of the scattering potential and is given in Eq.(\ref{4}).

Finally, we display the transfer matrix for the three-state model. It has the most compact form if we propagate the column vector $(\Psi_{l-1}^{(3)}, \Psi_{l}^{(2)}, \Psi_{l}^{(1)})^T$ with the shifted index for the third component. Then the transfer matrix is given by
\begin{equation}
\label{b6}
T_l=A^{-1}B \;,
\end{equation}
where the matrices $A$ and $B$ are written below :
\begin{equation}
\label{b7}
A=\left(
\begin{array}{ccc}
-1/\sqrt{2} & (U_0+2\epsilon_{l+1} -E)/J & 0\\
-1/2 & 0 & (\epsilon_{l}+\epsilon_{l+2} -E)/J \\
0 & -1/\sqrt{2} & -1/2
\end{array}
\right) \;,
\end{equation}
and
\begin{equation}
B=\left(
\begin{array}{ccc}
1/\sqrt{2} & 0 & 0\\
1/2 & 0 & 0 \\
(E-U_1-\epsilon_l-\epsilon_{l+1})/J   &  1/\sqrt{2} & 1/2
\end{array}
\right) \;.
\end{equation}

\begin{figure}
\center
\includegraphics[width=6cm,clip]{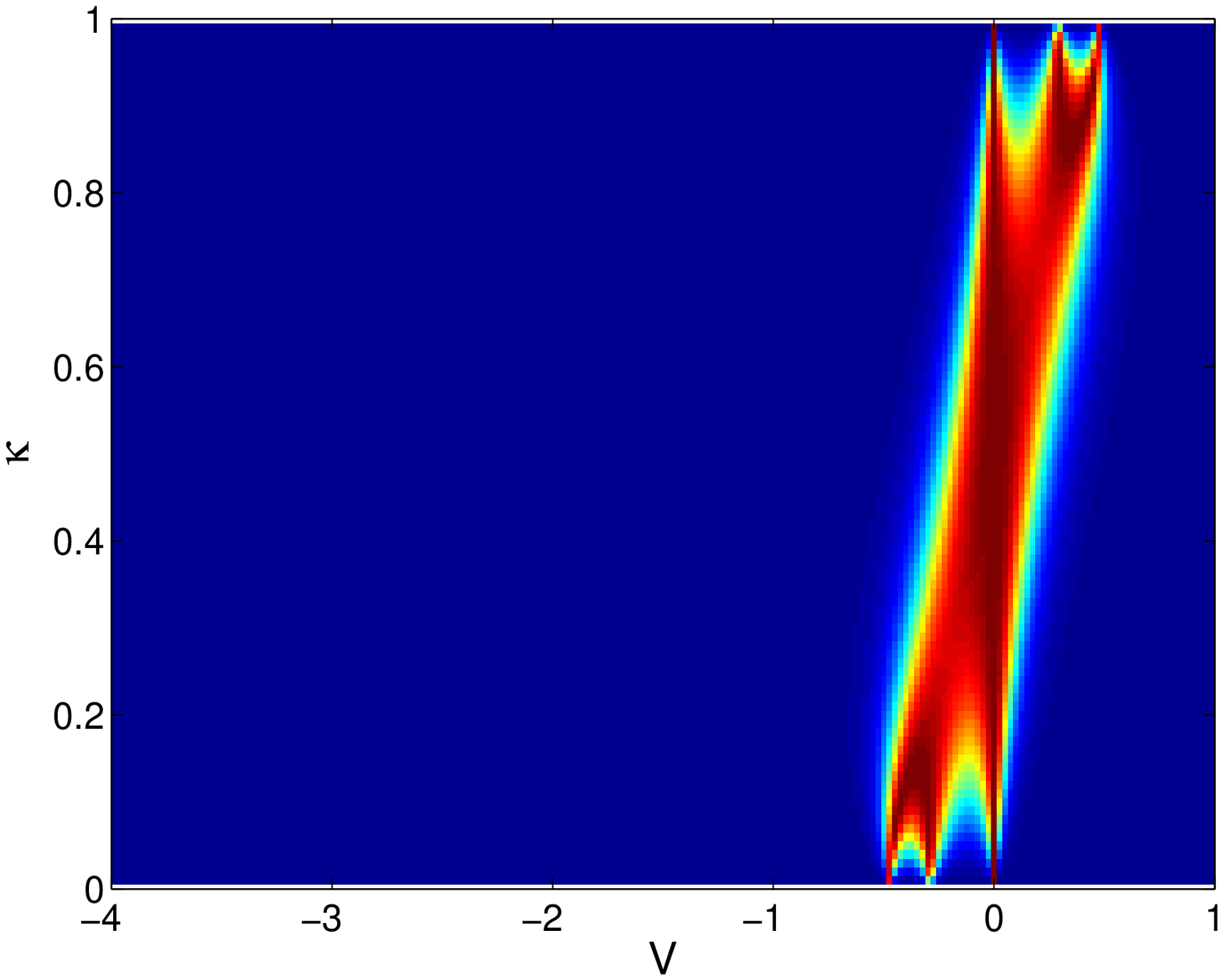}
\includegraphics[width=6cm,clip]{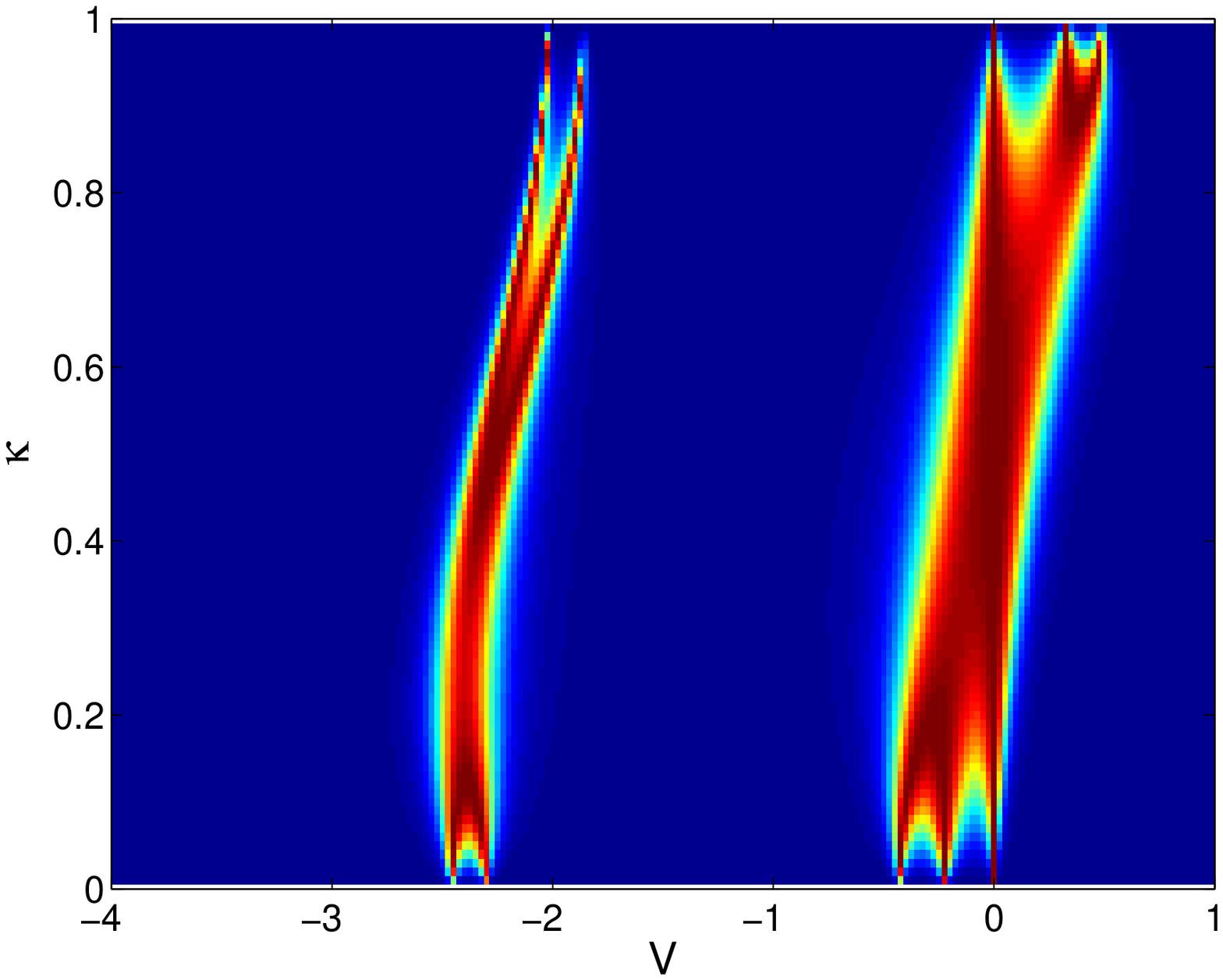}
\includegraphics[width=6cm,clip]{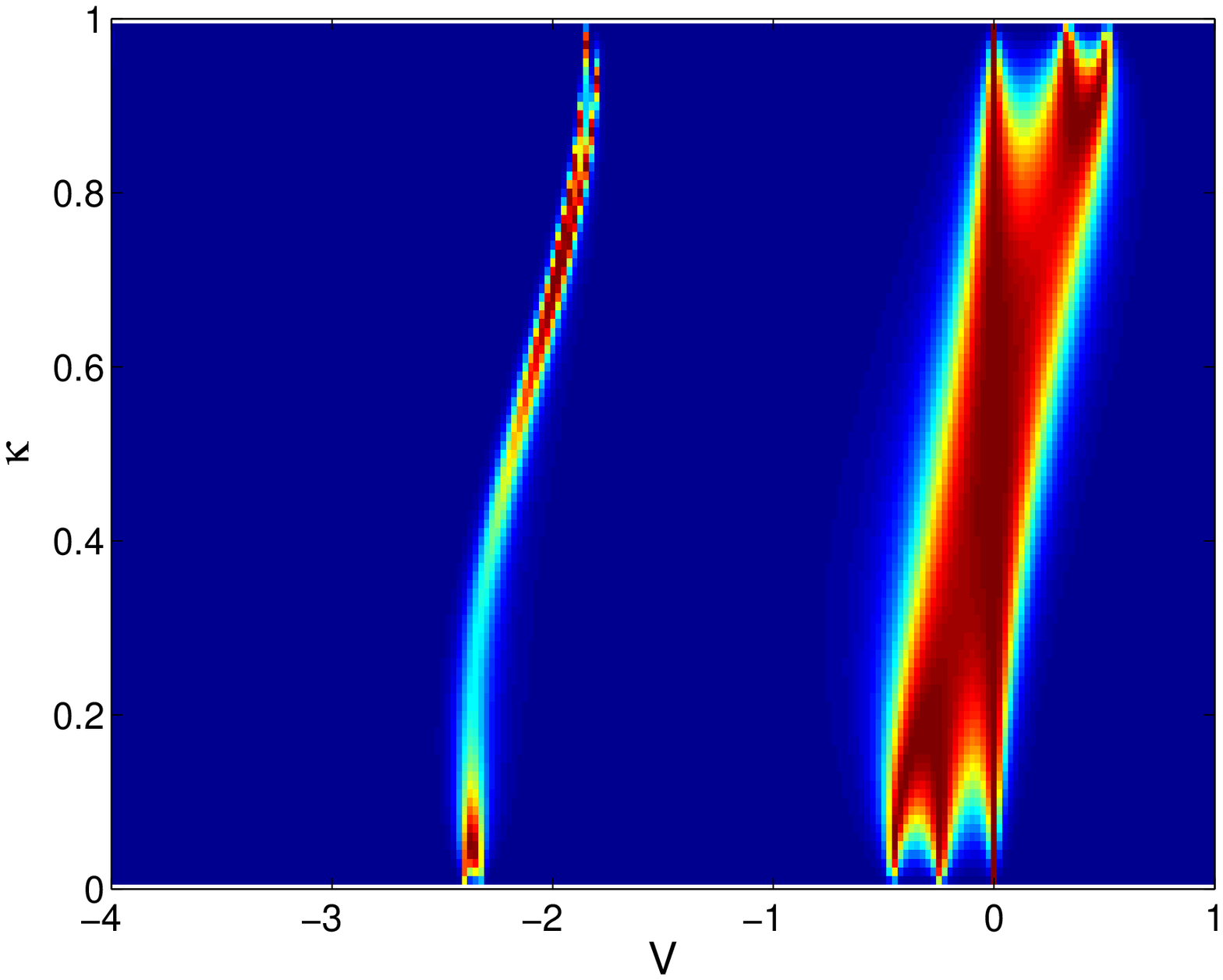}
\includegraphics[width=6cm,clip]{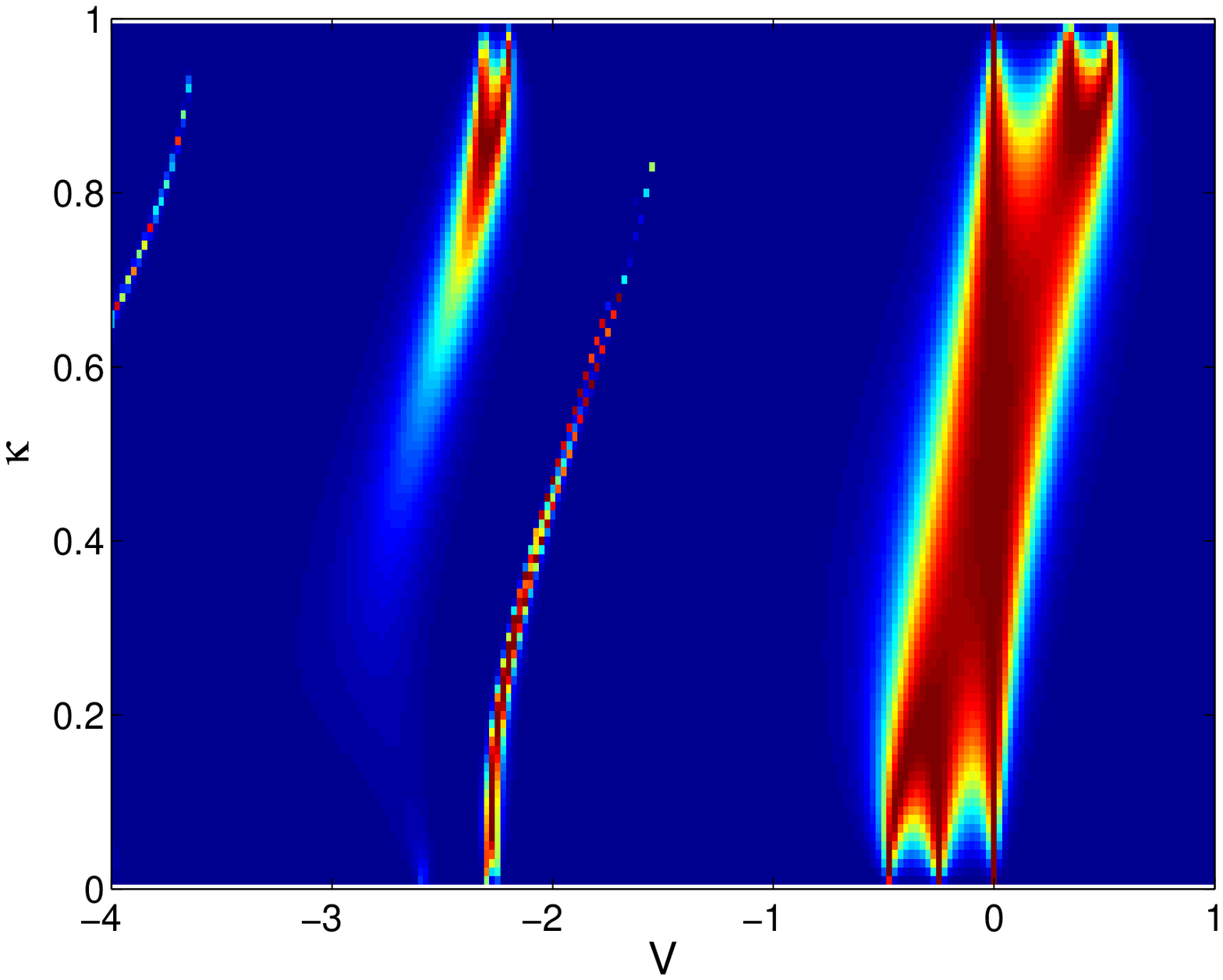}
\caption{Tunneling probability as predicted by the one-state (upper left), two-state (upper right), and three-state (lower raw) models. The system parameters are $J=1$, $\Delta=2$, and $U_1=-2$ in the lower left panel and $U_1=0$ in the lower right panel, the width of the Gaussian scattering potential $\sigma=0.65$.}
\label{Gfig9}
\end{figure}

In the paper, when discussing the tunneling probability, we focussed on the particular case $\kappa=\pi/2$ where the bound pair has the maximal group velocity. It is interesting to compare results of the one-, two-, and three-state models for other values of the quasimomentum. This comparison is given in Fig.~\ref{Gfig9} which shows the tunneling probability as the function the quasimomentum $\kappa$ and the amplitude $V$ of the external Gaussian potential ($\sigma=0.65$). The system parameters are $J=1$ and $U_0=U_1-2$,  which implies $\Delta=2$ in the two-state model and $I\approx 0.5$ in the one-state model. (In Fig.~\ref{Gfig9}(a) we used $I= 0.7321/2$, which is one half of the actual band width.)  A narrow window of the resonant tunneling (as predicted by the two-state model) and partial suppression of this resonant tunneling (as predicted by the three-state model) are clearly seen in the figure.



\begin{thebibliography}{99}
\bibitem{remark3}
We mention papers \cite{Kudo09,Krim11, Albr11}, which are also devoted to wave-packet dynamics of the bound pair.
\end{thebibliography}
\end{document}